%% file: paper.tex
\documentclass[12pt, draftclsnofoot, onecolumn]{IEEEtran}

\usepackage{tikz}
\usetikzlibrary{calc}
\usepackage{cite}
\usepackage{graphicx}
\usepackage{algorithm}
\usepackage[noend]{algorithmic}
\usepackage{epsfig}
\usepackage{epstopdf}
\usepackage{amssymb,amsmath}
\usepackage{amsmath}
\usepackage{amsthm}
\usepackage{amsfonts}
\usepackage{subfigure}
\usepackage{psfrag}
\usepackage{rotating}
\usepackage{latexsym}
\usepackage{dsfont}
\usepackage{stmaryrd}
\usepackage{amsthm}
\usepackage{amssymb}
\usepackage{amsmath}

\usepackage[force,almostfull]{textcomp}
\usepackage{fancyvrb}
\usepackage{textcomp}
\usepackage{url}
\usepackage{ifdraft}
\usepackage{url}
\usepackage{multirow}
\usepackage{hyperref}
\usepackage{rotating}
\usepackage{xspace}
\usepackage{array}
\usepackage{flushend}

\usepackage{enumerate}

\usepackage[utf8]{inputenc}
\newlength\figureheight
\newlength\figurewidth
\usepackage{graphics}
\usepackage{pgfplots}
\usepackage{siunitx}
\pgfplotsset{compat=newest}
\pgfplotsset{plot coordinates/math parser=false}

\usepackage{scalefnt}

\newtheoremstyle{specialcasestyle}{1mm}{1mm}{\upshape}{}{\bfseries\upshape}{.}{0mm}{}
\theoremstyle{specialcasestyle}

\newcommand{\var}{\operatorname{Var}}

\newcommand{\bb}[1]{\mathbb{#1}}

\newcommand{\gvn}{\mid}

\newcommand{\wnrv}{\mathop{\mathrm{WNRV}}}
\newcommand{\re}{\mathop{\mathrm{RE}}}

\renewcommand{\v}[1]{\boldsymbol{#1}} 
\newcommand{\di}{\mathrm{d}} 
\renewcommand{\c}[1]{\mathcal{#1}} 

\begin{document}

\title{A Universal Splitting Estimator for the Performance Evaluation of Wireless Communications Systems}

\author{Nadhir Ben Rached$^{1}$, Daniel MacKinlay$^2$,  Zdravko Botev$^2$, Ra\'ul Tempone$^{3,4}$, and Mohamed-Slim Alouini $^3$
\\
\thanks{\vspace{-0.2in}\hrule \vspace{0.2cm}

This work was supported by the KAUST Office of Sponsored Research (OSR) under Award No. URF/1/2584-01-01 and the Alexander von Humboldt Foundation.

$^1$ Chair of Mathematics for Uncertainty Quantification, Department of Mathematics, RWTH Aachen University, 52062 Aachen, Germany.

$^2$ School of Mathematics and Statistics, University of New South Wales (UNSW Sydney), 2052 Sydney, Australia.

$^3$ Computer, Electrical and Mathematical Sciences \& Engineering Division (CEMSE), King Abdullah University of Science and Technology (KAUST),  23955-6900 Thuwal, Saudi Arabia.

$^4$ Alexander von Humboldt Professor in Mathematics for Uncertainty Quantification, RWTH Aachen University, 52062 Aachen, Germany.

}
}
\date{}
\maketitle
\thispagestyle{empty}
\begin{abstract}
We propose  a unified rare-event estimator for the performance evaluation of wireless communication systems.
The estimator is derived from  the well-known multilevel splitting algorithm.
In its original form, the splitting algorithm cannot be applied to the  simulation and estimation of time-independent problems, because  splitting requires an underlying continuous-time Markov process whose trajectories can be split.
We tackle this problem by embedding the static problem of interest within a continuous-time Markov process, so that the target time-independent distribution becomes the distribution of the Markov process at a given time instant.
The main feature of the proposed  multilevel splitting algorithm is its large scope of applicability.
For illustration, we show how the same algorithm can be applied to the problem of estimating the cumulative distribution function (CDF) of sums of random variables (RVs), the CDF of partial sums of ordered RVs, the CDF of ratios of RVs, and the CDF of weighted sums of  Poisson RVs.
We investigate the computational efficiency of the proposed estimator via a number of simulation studies and find that it compares favorably with  existing estimators.
\end{abstract}

\begin{IEEEkeywords}
Rare event, performance evaluation, multilevel splitting algorithm, variance reduction
\end{IEEEkeywords}
\section{Introduction}
When assessing the performances of wireless communications systems operating over fading channels,  we often encounter sums of random variables (RVs)~\cite{alouini}.
For instance, the signal-to-noise-ratio (SNR) expression at the output of maximum ratio combining (MRC) is expressed as a sum of fading channel gains.
The sum of fading channel amplitudes is encountered when equal gain combining (EGC) diversity technique is employed.
A further challenging problem involving sums of RVs is the SNR expression at the output of generalised selection combining (GSC) receivers combined with MRC or EGC.
In this case, the SNR is a function of partial sums of ordered  fading channel gains (respectively fading channel amplitudes) when GSC is combined with MRC (respectively with EGC)~\cite{8472928}.
Many metrics are used to assess the performance of wireless communication systems.
Among these we distinguish the outage probability (OP), which is defined as the probability that the output SNR is less than a given threshold.
Computing the OP when diversity techniques are employed is equivalent to evaluating the cumulative distribution function (CDF) of sums of RVs:
sums of channel gains for  MRC, sums of channel amplitudes for EGC, and partial sums of ordered channel gains for GSC/MRC or ordered channels amplitudes for GSC/EGC, see~\cite{7328688,8472928}.
Sums of RVs are also encountered in the evaluation of the OP in the presence of co-channel interferences and noise.
The signal-to-interference-plus-noise ratio (SINR) is expressed in this case as a ratio of the desired power signal and the sum of interfering power signals plus noise.
Thus, in this case the OP computation is equivalent to evaluating the complementary CDF of sums of RVs \cite{7857009,botev2019fast}.

In addition to the previously mentioned applications, in which the considered RVs are typically continuous, sums of discrete RVs  also involve challenging computations in  wireless communication applications.
Of particular interest is the probability that a weighted sum of independent Poisson RVs falls below  a small threshold, which represents the probability of missed detection  of pointing and alignment of free space optical communication systems \cite{Bashir}.
In addition to that,  weighted sums of independent Poisson RVs are also encountered  in reliability and physics
\cite{scaled_poisson}. It is important to mention that all the above applications can be grouped into a single framework. In fact, they can be reduced to the problem of investigating the probability that a quasi-monotone functional (such as weighted sums and ratios of RVs) falls below a threshold (a formal definition will be given in the core of the paper).

In this article, we propose to tackle problems  under the quasi-monotonicity framework using a novel dynamic multilevel splitting method.
In its original form \cite{kahn1951estimation}, the multilevel splitting algorithm evolves a continuous-time Markov process whose trajectories are  split in an attempt to help the process enter a small set of interest. The classical splitting method cannot be applied to an estimation problem with static or time-independent distributions. 
We solve this problem by embedding the static distribution of interest  within a continuous-time Markov process.  In this way  the Markov process has exactly the target distribution at a specified time instant.
With this embedding in hand, we can apply any suitable version of the dynamic splitting algorithm to estimate many performance metrics. The main contributions of the present work are summarized as follows:

\begin{itemize}
\item The proposed algorithm presents a unified approach to tackling all of the above-mentioned problems with a single algorithm, the sole assumption is the quasi-monotonicity which will be defined later. The estimation of the CDF of the sum of independent RVs is the first application of the proposed algorithm. Interestingly, we do not impose any assumptions about the distributions of the RVs in the summation.  
    This unified feature together with its implementation simplicity are the main advantages over existing estimators.
\item A second application of the proposed approach is to efficiently estimate OP values in the presence of co-channel interferences and noise.
    Again,  a unifying aspect of our splitting method is that we do not need to impose any assumptions about the nature of the distributions of the desired and interfering powers.
\item We further our analysis by applying the modified dynamic splitting algorithm to the problem of estimating the CDF of partial sums of ordered RVs.
    This serves to compute OP values at the output of GSC receivers combined with either MRC or EGC.
    We compare the proposed estimator with existing importance sampling and conditional Monte Carlo (MC) estimators \cite{8472928}.
\item A close look at the literature reveals that the existing estimators of the CDF of sums of RVs have been constructed under the assumption of continuous distributions.
    An important advantage of our method is its ability to deal with sums of  discrete RVs as well. As an example, we  apply the proposed method to the calculation of tail probabilities of weighted sums of independent Poisson RVs.
\end{itemize}

It is worth mentioning that we do not claim that the proposed splitting algorithm yields the most efficient estimator for the problems we consider in this paper (in some settings, there may already be a particular estimator that is tailor-made for a particular problem and hence exhibits better performances).
Rather, we  emphasize the wide scope of applicability of the algorithm and its implementation simplicity. 
It is also important to note that, despite recent advances in the theory and practice of the splitting algorithm for rare-event simulation,  its popularity  in the field of wireless communications is still limited.
Therefore, this work represents a relevant step in promoting the use of this algorithm for quickly and accurately assessing the performance of wireless communications systems in view of the great interest in ultra-reliable wireless data transfer \cite{7041045,6g}. More specifically, for ultra-reliable 5G or 6G systems, one often encounter error probabilities (such that the bit error rate or the outage probability) with very small values, say of the order of $10^{-9}$. Thus, proposing an algorithm that accurately and quickly estimate such  rare event probabilities would be of paramount practical interest.

The rest of the paper is organized as follows.
In Section II, we present the modified version of the multilevel splitting algorithm that deals with static problems under the so-called quasi-monotonicity assumption.
In  Section III, we demonstrate the wide scope of applicability of the proposed algorithm by applying it to four different challenging problems. In section IV, two heuristics for choosing optimal splitting levels are described.
In Section V, we give a number of illustrative simulation results and compare them with some of the existing estimators and benchmarks.

\section{Splitting and Embedding via Gamma Subordinator}
In this section we introduce a novel  method for estimating a rare-event probability of the form 
\begin{equation}
\label{ell}
\ell=\bb P[S(\v X)\leq \gamma],
\end{equation}
where $\v X:=(X_1,\ldots,X_n)^\top$ is a random vector of independent RVs with cdf $F_i(x)=\bb P[X_i\leq x]$, and
 the so-called \emph{importance function} $S:\bb R^n\mapsto \bb R$ is a \emph{quasi-monotone}  function, defined as follows. Suppose that we divide the index set $\{1,\ldots,n\}$ into two sets $\c I$ and $\c D$, where either one of them can  be empty.   $S$ is said to be  a quasi-monotone function
if  $x_i\leq y_i,\;i\in \c I$ and $x_i\geq  y_i,\;i\in \c D$, implies $S(\v x)\leq S(\v y)$. For example,
$S(\v x)=x_1+\cdots+x_n$ satisfies the \emph{quasi-monotonicity} property with $\c I=\{1,\ldots,n \}$.

 It turns out that a  number of useful models either possess this quasi-monotonicity property, or can be transformed into  models possessing it (see, for example, \cite{lam2015dynamic}).
The proposed algorithm can estimate $\ell$ under the quasi-monotonicity assumption. It is a  variant of the \emph{classical dynamic splitting} algorithm for  simulation of  a Markov processes conditional on a rare event 
\cite{kahn1951estimation}. In classical splitting  the state space of the Markov process is decomposed into nested subsets so that the rare event is represented as the intersection of decreasing events. Within each subset the sample paths of the Markov process are split into multiple copies with the goal of promoting more occurrences of the rare event. 

In its original form, classical splitting cannot be applied to estimate \eqref{ell}, because there is no underlying Markov process that we can split.  Our proposed splitting algorithm can be viewed as
an ingenious way of transforming the static performance metric \eqref{ell} so that it involves the simulation of a continuous-time Markov process. The  idea is to embed the
density of $\v X$  within a continuous-time Markov process $\{\v X(t), t\geq 0\}$ in such a way that $\v X(1)$ has the same distribution as $\v X$. 
Given this embedding, we can then apply the original splitting idea of \cite{kahn1951estimation,lam2015dynamic}.


\begin{figure}[htb]
    \centering
    \includegraphics[width=\linewidth]{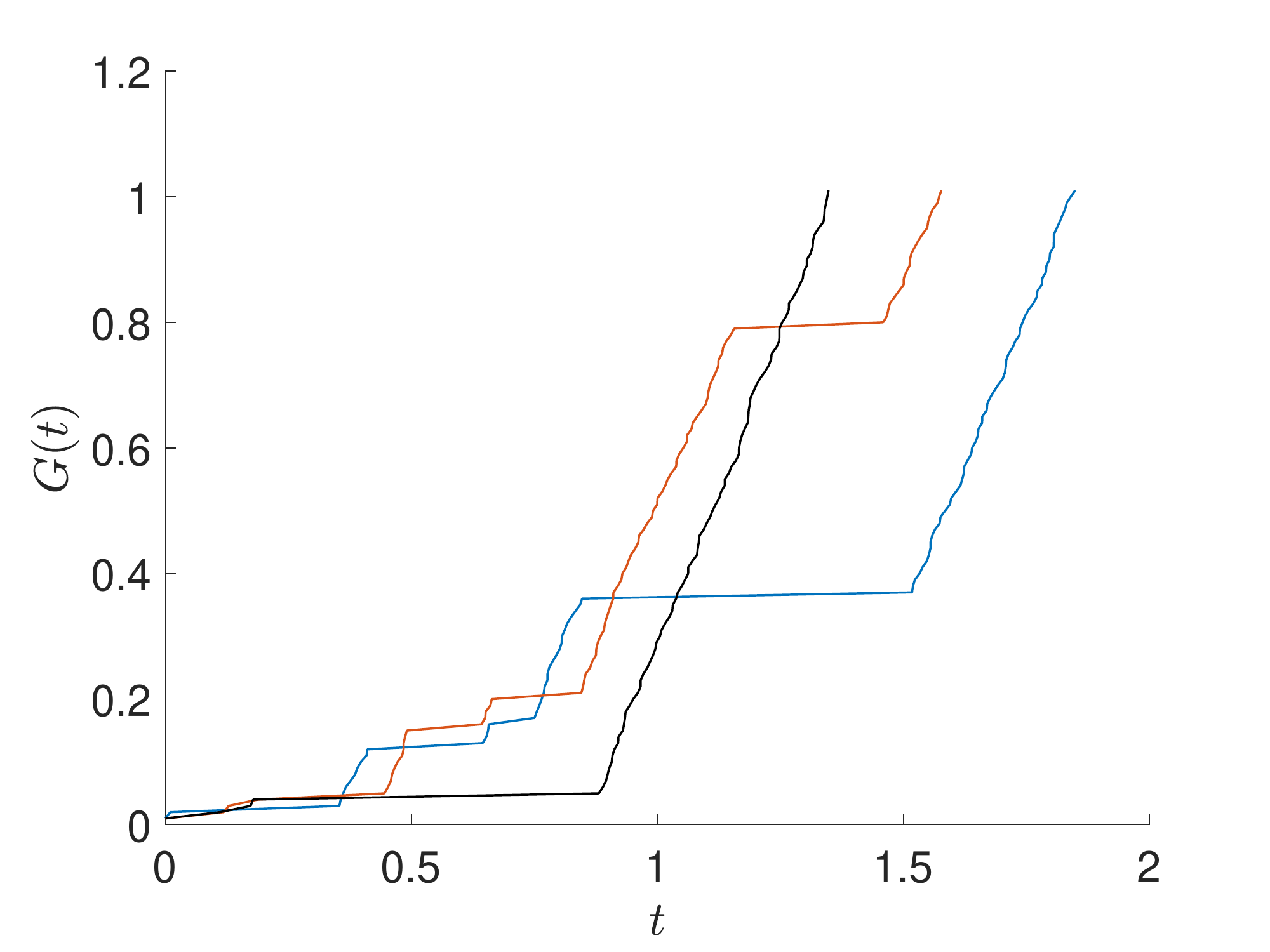}
    \caption{Three independent realizations of a Gamma process.}
\end{figure}

\subsection{Gamma Process}
A suitable  Markov process for our purposes is the multivariate Gamma process, defined as follows. 
A Gamma process $\{G(t),t\geq 0\}$ with $G(0)=0$ is a continuous-time  and continuous state space  process with the following properties:
(1) the increments of $\{G(t),t\geq 0\}$ are stationary and Gamma distributed, that is, $G(t+s)-G(t)$ has the same distribution\footnote{Here $\mathsf{Gamma}(\alpha,\beta)$ be the Gamma distribution with density $\beta^\alpha x^{\alpha-1}\exp(-\beta x)/\Gamma(\alpha)$.} as $G(s)\sim\mathsf{Gamma}(s,1)$ for $t,s>0$; (2)
the increments are independent, that is,
 $G(t_i)-G(t_{i-1})$ for $i=1,2,\ldots$ are independent for 
$0\leq t_0<t_1<\cdots$ and (3) we have $\lim_{s\downarrow 0}\bb P[|G(t+s)-G(t)|\geq\epsilon]=0$. 
Fig. 1 above shows the sample paths of three independent Gamma processes.

We also recall the definition of a multivariate Gamma process.
Let $\{G_i(t),t\geq 0\}$ for $i=1,\ldots,n$ be $n$ independent Gamma processes. Then, 
$\v G(t)=(G_1(t),\ldots,G_n(t))^\top$ is a multivariate Gamma process, which can be simulated at given times $0=t_0<t_1<t_2<\cdots<t_L=1$, as follows. 
\begin{algorithm}[H]
\caption{Simulating $\{\v G(t),t\geq 0\}$ at $\{t_i\}_{i=1}^L$}
\begin{algorithmic}
\REQUIRE {Sequence of times  $\{t_i\}_{i=1}^L$}
\STATE{$\v G(t_0)\leftarrow \v 0$}
\FOR{$i=1,\ldots,L$}
\FOR{$k=1,\ldots,n$}
\STATE {Simulate $G^*_k\sim \mathsf{Gamma}(t_i-t_{i-1},1)$,  independently.}
\ENDFOR
\STATE{$\v G^*(t_i-t_{i-1})\leftarrow (G^*_1,\ldots,G^*_n)^\top$}
\STATE{ $\v G(t_{i})\leftarrow \v G(t_{i-1})+\v G^*(t_i-t_{i-1})$}
\ENDFOR
\RETURN{$\v G(t_{1}),\ldots,\v G(t_{L})$}
\end{algorithmic}
\end{algorithm}

\subsection{Embedding with Gamma Process}

 With the Gamma process in hand, 
we proceed to embed the 
 distribution of $\v X=(X_1,\ldots,X_n)^\top$ in \eqref{ell} within a continuous-time Markov process as follows. Let $\{\v G(t), t\geq 0\}$ 
be a multivariate Gamma process.  Define the vector $\v X(t)=(X_1(t),\ldots,X_n(t))^\top$ through the random variables
\begin{equation}
\label{mapto gamma}
\begin{split}
X_i(t)&=F^{-1}_i(1-\exp(-G_i(t))),\quad i\in \c I\\ 
X_i(t)&=F^{-1}_i(\exp(-G_i(t))),\quad i\in \c D.
\end{split}
\end{equation}
Then, we have the realizable  almost sure partial (and in fact total) ordering for the vectors for all 
 $s,t>0$: 
\[
\begin{split}
X_i(t)&\leq X_i(t+s),\;i\in\c I \\
X_i(t)&\geq X_i(t+s),\;i\in\c D.
\end{split}
\]  More importantly, at time $t=1$ each $X_i(1)$ has the 
desired distribution $\bb P[X_i(1)\leq x]=F_i(x)$. In this way, we 
view the outcome of the original vector $\v X$ as a snapshot of the state of a 
multivariate continuous-time process $\{\v X(t),t\geq 0\}$ at the instant $t=1$. 

Note that the only thing that we change from one estimation problem to the next is 
the inverse CDF formulas in the transformation \eqref{mapto gamma}.
It is in this sense that the method in the next section has a wide scope of applicability. 

\subsection{Main Splitting Algorithm}
The above embedding now has a number of consequences.
First,  $S(\v X(t))$ is an increasing function of $t$ and this  implies that 
a unique exit time exists
\begin{equation*}
\label{exit time}
\tau_\gamma=\sup\{t:S(\v X(t))\leq\gamma\}
\end{equation*}
and in fact
$
\bb P[S(\v X(t))\leq \gamma]=\bb P[\tau_\gamma>t]\,.
$
Second, if
\[
0=t_0<t_1<\cdots<t_L=1
\]
is any sequence of increasing times, the state space can be decomposed into the decreasing sequence of  events:
 \[
 \{S(\v X(t_0))\leq \gamma\}\supseteq \{S(\v X(t_1))\leq \gamma\}\cdots\supseteq \{S(\v X(t_{L}))\leq \gamma\}.
\] 
Therefore, we have the following decomposition of \eqref{ell}
\[
\begin{split}
\ell&=\bb P[S(\v X(1))\leq \gamma]\\
&=\prod_{l=1}^L\bb P[S(\v X(t_{l}))\leq \gamma \gvn S(\v X(t_{l-1}))\leq \gamma]\\
&=\prod_{l=1}^L\bb P[\tau_{\gamma}>t_l\gvn \tau_{\gamma}>t_{l-1}]\,.
\end{split}
\]

\begin{figure}[htb]
    \centering
    \def\svgwidth{\columnwidth}   
    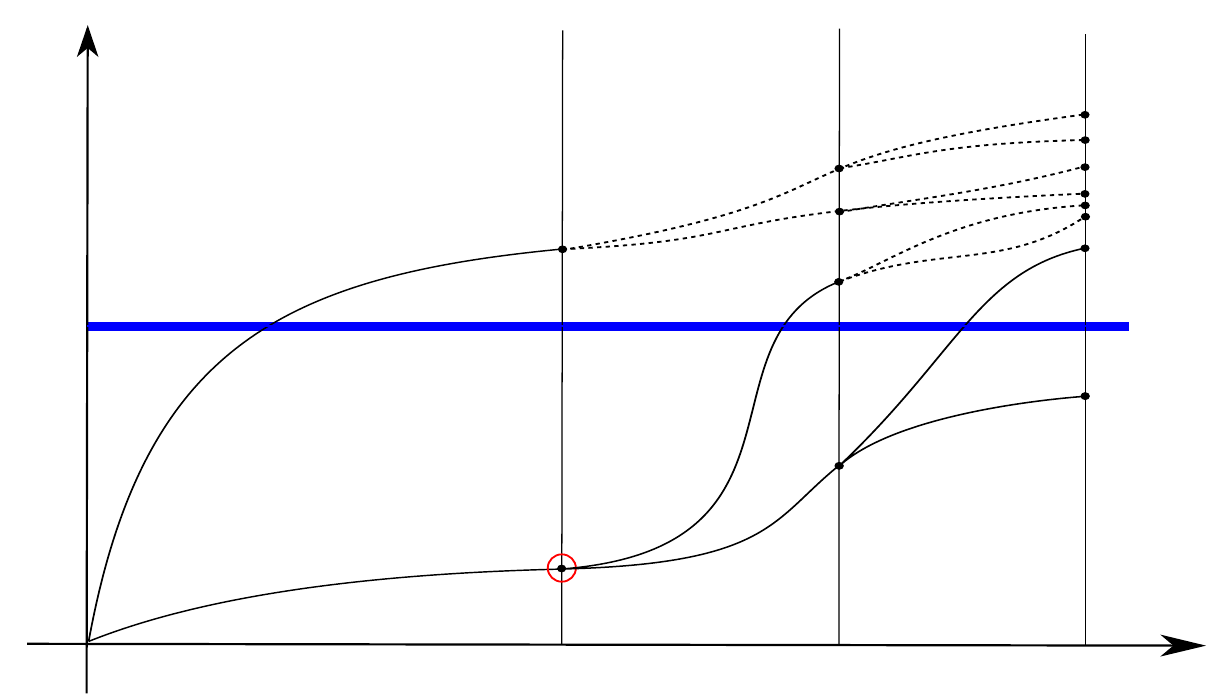
    \caption{Stylized representation of the splitting process.
		Splitting factor is two
		The instant of splitting is denoted with red circles. The dotted trajectories that end up above the $\gamma$ threshold are not simulated. Only the paths corresponding to the solid lines are simulated.}\label{splitdiag}
\end{figure}

This decomposition now suggests that, given $\{t_l\}_{l=1}^L$, we may estimate the
conditional probabilities $\bb P[\tau_{\gamma}>t_l\gvn \tau_{\gamma}>t_{l-1}]$ sequentially, just as in the classical and generalized splitting algorithms \cite{kahn1951estimation,l2018generalized}. Initially, we simulate
$s$ number of independent versions of $\v X(t_1)$. Then, we collect all the $\v X(t_1)$'s for which  $S(\v X(t_1))\leq\gamma$ (that is, $\tau_\gamma>t_1$) in the set $\c X_1$. Of course, an estimate of
 $\bb P[\tau_\gamma>t_1\gvn\tau_\gamma>t_0]$ is $|\c X_1|/s$. Next,
we randomly resample $s$ Markov states from $\c X_1$. Each such Markov state $\v X(t_1)$
is now evolved for an extra $t_2-t_1$ period of time to obtain
the new state $\v X(t_2)$. Again,  we collect all the $\v X(t_2)$'s for which  $S(\v X(t_2))\leq\gamma$  in the set $\c X_2$, and estimate
$\bb P[\tau_\gamma>t_2\gvn\tau_\gamma>t_1]$ via $|\c X_2|/s$.
We repeat this process until we get the final time/level $t_L$
and then take the product of all the conditional probability estimates to obtain:
\[
\hat \ell=\prod_{l=1}^L \frac{|\c X_l|}{s}.
\]
Fig. \ref{splitdiag} depicts the typical evolution of  trajectories of the Markov process $\{\v X(t),t\geq 0\}$ in the case where $s=2$ and $L=3$.
Observe how, as a result of the quasi-monotonicity property, the paths/trajectories of the Markov process are monotonically increasing. 
Since there is only one trajectory that survives till time $t_L=1$, the estimate here is thus $\hat\ell=1/2^3=1/8$. 

In summary, the main splitting algorithm written in pseudo-code is as follows. 

\begin{algorithm}[H]
\caption{Splitting for estimating $\ell$}
\label{alg:split}
\begin{algorithmic}[1]
\REQUIRE {Sample size $s$ and intermediate  levels $\{t_i\}_{i=0}^L$}
\STATE{$\v G(t_0)\leftarrow \v 0$}
\STATE{$\c X_{0}\leftarrow \{\v G(t_0),\ldots,\v G(t_0)\}$, a set of $s$ copies of $\v G(t_0)$}
 \FOR{$i=1,\ldots,L$}
     \FOR {$k=1,\ldots,s$}
		\STATE{Let $\v G(t_{i-1})$ be a randomly chosen member of $\c X_{i-1}$.}
   
			\STATE{
Simulate	$\v G^*(t_i-t_{i-1})\leftarrow (G_1^*(t_1),\ldots,G_{n}^*(t_1))^\top$
			}
			\STATE{ $\v G(t_{i})\leftarrow \v G(t_{i-1})+\v G^*(t_i-t_{i-1})$}
			\STATE{Compute $\v X(t_i)$ from $\v G(t_{i})$ via \eqref{mapto gamma}}
       \IF{$ S(\v X(t_i)) \leq \gamma$}
         \STATE{add $\v G(t_i)$ to $\c X_i$}
       \ENDIF 
   \ENDFOR
 \ENDFOR
 \RETURN {$\hat\ell \leftarrow \frac{1}{s^L}\prod_{i=1}^L |\c X_i|$ as an estimate of $\ell$.}
\end{algorithmic}
\end{algorithm}
We now discuss how this same algorithm applies to a range of different estimation problems. 

\section{Applications}
\subsection{CDF of the Sum of Independent RVs}
The CDF of the sum of independent RVs serves to compute OP at the output of EGC and MRC receivers.
In both cases, the OP could be expressed as \eqref{ell},
where $S(\boldsymbol{x})=\sum_{i=1}^{n}{x_i}$ and the RVs $X_1,\cdots,X_n$  are assumed independent and represent the fading channel amplitudes for EGC receivers and the fading channel gains for MRC receivers (see, for example, \cite{7328688}).
In most of the cases, the CDF of the sum of independent RVs is not known in a closed-form expression.
This includes the CDF of sums of Log-normal, Rayleigh, Nakagami, Rice, etc.
The calculation of $\ell$ has been extensively investigated  in the literature through either approximation methods \cite{1388730,4275022,1275712,4814351}
or efficient simulation techniques
\cite{botev2019fast,Nadhir_SLN,7328688,8647776,7835220,laub2019monte}.

The application of the dynamic multilevel splitting algorithm to this setting is straightforward, because $S(\v x)$  satisfies the quasi-monotonicity assumption.
Note that in order to be able to  employ the proposed algorithm, the inverse of the CDF of each of $X_i$, $i=1,\cdots,n$, has to be known in closed-form, or at least be efficient to calculate.
This assumption does not introduce a serious limitation since for most of the useful distributions the inverse CDF can be computed analytically (for example, the Log-Normal, and the Generalized Gamma, which includes the Gamma, the Nakagami-m, and the Weibull distributions).


\subsection{CDF of the Ratio of Independent Positive RVs}
The second application that we  investigate is the evaluation of
the OP values in the presence of co-channel interferences and noise
\begin{align}
\ell= \bb P \left [ \frac{X_1}{\sum_{i=2}^{n}{X_i}+\eta} \leq \gamma\right ].
\end{align}
The RV $X_1$ represents the received power of the useful signal,
$X_2,\cdots,X_n$ are the received  powers of the $n-1$ interfering sources and
$\eta>0$ is the variance of the additive white Gaussian noise.
The RVs $X_1,X_2,\cdots,X_n$ are assumed to be independent and non-negative.
None of the currently existing  approximation approaches (e.g., \cite{7006713,6661325,7769235}) or simulation approaches \cite{7857009,botev2017accurate,8462177} are generic and easily applicable for a range of possible distributions.
Thus, constructing a simple generic estimator that works for a wide range of distributions is of practical interest.

Again, this estimation problem fits our  multilevel splitting framework, because
\[
S(\boldsymbol{x}):=
\frac{x_1}{\sum_{i=2}^{n}{x_i}+\eta} 
\]
is quasi-monotone in
$\boldsymbol{x}=(x_1,x_2,\cdots,x_n)^\top$, provided that we chose 
$\c I=\{1\}$ and $\c D=\{2,\ldots,n\}$ in \eqref{mapto gamma}.

\subsection{CDF of the Sum of Order Statistics of Independent RVs}
A further practical problem is the computation of the CDF of sums of order statistics of independent RVs.
Such a quantity is a useful metric as it corresponds to the OP at the output of GSC receivers combined with either MRC or EGC diversity techniques. In other words, we need to compute \eqref{ell} with
\[
S(\v X ):=\sum_{i=1}^{\bar{n}}{X^{(i)}},
\]
with $1 \leq \bar{n} \leq n$ and $X^{(1)} \geq X^{(2)}\geq  \cdots \geq X^{(n)}$.
Note that $X^{(1)}, \cdots, X^{(\bar{n})}$ represent the best $\bar{n}$ out of
$n$ ordered fading channel amplitudes for GSC/EGC and fading channel gains for GSC/MRC.
Few works addressing this problem exist in the literature.
Closed-form results are available when $X_1,\cdots,X_n$ are exponential or Gamma distributed RVs \cite{5605378,7953495}.
Specialized simulation methods have also been proposed in some recent works \cite{8472928,8462177}, 
 in which the RVs $X_1,\cdots,X_n$  are distributed according to
the Generalised Gamma and the Log-normal variates.

The application of the proposed dynamic multilevel splitting algorithm to the current problem is as follows.
Given that $S(\boldsymbol{x})= \sum_{i=1}^{\bar{n}}{x^{(i)}}$, $\boldsymbol{x}=(x_1,\cdots,x_n)^\top$, satisfies the quasi-monotonicity, it suffices then to embed each of the $X_i$, $i=1,\cdots,n$ with  a monotonically increasing  transformation (the first transformation in (\ref{mapto gamma})). Again,  we do not assume any particular  distribution for the RVs $X_i$, $i=1,\cdots,n$.

\subsection{CDF of the Weighted Sum of Independent Poisson RVs}

Another advantage of the proposed splitting algorithm is its ability to deal with the CDF of  sums of discrete RVs just as easily as with continuous RVs.
Let $X_1,X_2,\cdots,X_n$ be a sequence of independent Poisson RVs with density
\begin{align*}
\bb P[X_i=k]=\lambda_i^k \exp \left ( -\lambda_i \right )/k!, \quad k=0,1,2,\ldots
\end{align*}
Our interest is to efficiently estimate the CDF of weighted sum of the $X_i$'s. In other words, estimate \eqref{ell} with
\[
S(\v x):=\sum_{i=1}^{n}{w_ix_i},
\]
where $w_i$ are non-negative weights.
\subsubsection{Splitting Estimator}
For this problem,
it is not necessary to use the Gamma  process embedding (as in Algorithm~\ref{alg:split}) in order to apply the dynamic (time-evolving) splitting algorithm. This is because, each Poisson variable $X_i$ already  has the same distribution as the continuous-time Markov-jump (Poisson) process $\{X_i(t),t\geq 0\}$ with $X_i(0)=0$ and density
\begin{align*}
\bb P[X_i(t)=k]=(\lambda_i t)^k \exp \left ( -\lambda_i t \right )/k!, \quad k=0,1,2,\ldots
\end{align*}
In fact, the Poisson process shares the same key properties as the Gamma process: 
 (1) it has independent increments; (2) $X_i(t+s)-X_i(t)$ has a Poisson distribution with rate $s \lambda$ for $s,t \geq 0$ and (3) $\lim_{s\downarrow 0}\bb P[|X_i(t+s)-X_i(t)|\geq \epsilon]=0$ for any $\epsilon>0$.

Thus, without the need for any embedding, we have $\bb P[X_i(1)=k]=\bb P[X_i=k]$ for all $k$. Algorithm~\ref{alg:split} remains the same, except that we replace all instances of $\v G$ with $\v X$ and remove line 8. 

\subsubsection{Importance Sampling Estimator}
To the best of the authors' knowledge, a MC method that estimates the CDF of weighted sums of Poisson RVs does not exist.
Therefore, constructing a competing estimator to the multilevel splitting estimator is of practical interest.
Our second estimator is based on the use of the method of importance sampling \cite{botev2014variance}.
We consider an importance sampling density given by scaling the rate of each of the Poisson variates $X_i,\cdots, X_n$.
The  density of a scaled variate is
\begin{align}\label{is}
(\lambda_i\theta)^k \exp \left ( -\lambda_i\theta \right )/k!, \text{     }k=0,1,2,\cdots
\end{align}
with $0<\theta<1$ and goes to zero as $\gamma$ goes to zero.
Thus, the importance sampling estimator is given by
\begin{align*}
\hat{\ell}_\mathrm{IS}=\frac{1}{m}\sum_{i=1}^{m}
\textbf{1}_{\left \{ \sum_{j=1}^{n}w_jX_j(\omega_i)
\leq \gamma\right \}}\prod_{j=1}^{n}
\frac{\exp \left (-\lambda_i(1-\theta) \right )}{\theta^{X_j(\omega_i)}},
\end{align*}
where $m$ is the number of independent MC replications and for each replication, $\{X_j(\omega_i)\}_{j=1}^{n}$ are sampled independently according to \eqref{is}.
We choose the parameter $\theta$ such that the expected value of $\sum_{i=1}^{n}{w_iX_i}$ under \eqref{is}  is equal to $\gamma$.
A straightforward computation shows that the value of $\theta$ is given by
\begin{align}
\theta=\frac{\gamma}{\sum_{i=1}^{n}{w_i\lambda_i}}.
\end{align}

\section{Level Selection Heuristics}
In this section, we provide two heuristic algorithms for selecting  good intermediate times $t_1<t_2<\ldots<t_L$ and  the number of levels $L$  at which splitting is applied. The first one is universally applicable, and the second one applies to sums of random variables. The criterion that we adopt is that the splitting levels are adaptively chosen such that the following constraint is approximately satisfied
\begin{align}\label{level_criterion}
 \bb P[\tau_{\gamma}>t_l\gvn \tau_{\gamma}>t_{l-1}]=\frac{\bb P[\tau_{\gamma}>t_l]}{\bb P[\tau_{\gamma}>t_{l-1}]}=\bar{p}
\end{align}
for all $l=1,\cdots,L$ and for some fixed value $\bar{p}$ (chosen to be equal to $0.1$ in our numerical results).
\subsection{Inverse Complementary CDF Method}
First, we describe the inverse complementary CDF method for selecting levels that approximately satisfy \eqref{level_criterion}. In principle, this method can always be applied.
Let us denote by $\bar{F}(\cdot)$ the complementary CDF of the RV $\tau_{\gamma}$, then the optimal levels are selected to approximately satisfy
\begin{align}\label{opt_level}
t_l=\bar{F}^{-1}(\bar{p}^{l}), \text{  }  l=1,\cdots,L.
\end{align}
 for each $l=1,\cdots,L$.

\begin{figure}[H]
    \centering
    \includegraphics[width=\linewidth]{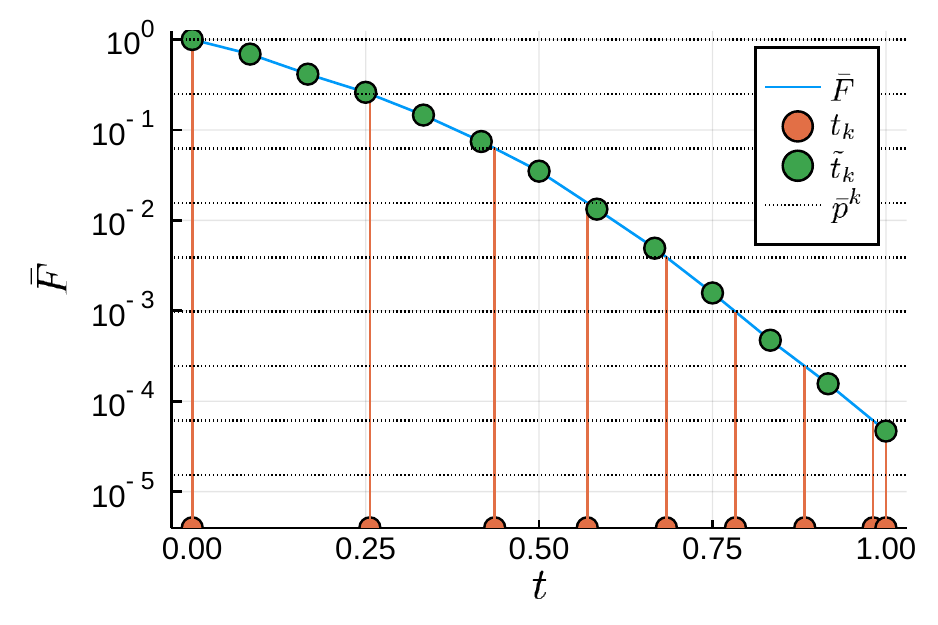}
    \caption{Graphical depiction of method of level selection.
		}
\end{figure}

The first step  is to run a crude pilot version of the proposed multilevel splitting algorithm using equally spaced intermediate levels $\tilde t_1,\cdots \tilde  t_L$. The output of this pilot run are the estimates  
\[
\widehat{\bar{F}}(\tilde t_l)=\prod_{i=1}^l |\c X_i|/s^l
\]
 for $l=1,\cdots,L$ (depicted as the green dots in Fig. 3 for each $\tilde t_l$).
Next, linear interpolation is used between the (green) points $(\tilde t_l,\widehat{\bar{F}}(\tilde t_l))$ to approximate the complementary CDF $\bar{F}(\cdot)$ on the interval $[0,1]$ (depicted as the blue curve on Fig. 3). Finally, the  levels that (approximately) satisfy \eqref{opt_level}  are computed by inverting the estimate of $\bar{F}(\cdot)$ (that is, the blue curve on Fig. 3) at the points $\bar p^l$. These final values are  depicted as  the red points on Fig. 3.

\subsection{Using Lower Bounds for Sums}
This method is not universally applicable. We explain below its scope of applicability. To begin with, assume that $S(\v x):=\sum_{i} x_i$. Then, to select good levels, we  use an analytical formula for the lower bound of $\bb P[S(\v X(t)) \leq \gamma]$
Specifically, since the term $\bb P \left[ \sum_{j=1}^{n}{X_j(t_1)} \leq \gamma \right ]$ is lower bounded by
$\bb P \left [ X_1(t_1) \leq \gamma/n,\cdots, X_n(t_1) \leq \gamma/n\right ]$,
we  select $t_1$ such that
\begin{align}
\nonumber &\bb P \left [ X_1(t_1) \leq \gamma/n,\cdots, X_n(t_1) \leq \gamma/n\right ]\\
\label{lower_bound}=&\prod_{i=1}^{n}\bb P \left [ X_i(t_1) \leq \gamma/n\right ]=\bar{p}.
\end{align}
Such an $t_1$ can be computed using a one-dimensional (off-the-shelf) root-finding algorithm, because:
\[
\begin{split}
\bb P \left [ X_i(t_1) \leq \gamma/n\right ]&=
\bb P[G(t_1)\leq -\ln \overline F_i(\gamma/n)]\\
&=\Gamma(-\ln \overline F_i(\gamma/n),t_1)/\Gamma(t_1),
\end{split}
\]
where 
\[
\Gamma(x,\alpha):=\int_0^x s^{\alpha-1}\exp(-s)\di s
\] is the
 lower incomplete Gamma function and $\Gamma(\alpha):=\Gamma(\infty,\alpha)$. 
Similarly,  we may select $t_i$ as the root  of the nonlinear equation
\begin{align*}
\bb P \left [ X_1(t_i) \leq \gamma/n,\cdots, X_n(t_i) \leq \gamma/n\right ]=(\bar{p})^i.
\end{align*}

Note that this heuristic can be easily adjusted to the case of the sum order statistics and to the weighted sum of Poisson RVs. 
In fact, in the problem of weighted sums of Poisson,  the lower bound becomes
 \begin{align}
 \ell=\bb P \left [ \sum_{i=1}^{n}{w_i X_i(t)} \leq \gamma \right  ] \geq \prod_{i=1}^{n}{\bb P \left [ X_i(t) \leq \frac{\gamma}{w_i n}\right ]}, 
 \end{align}
where each probability in the product is the CDF of a Poisson RV with rate $\lambda_i t$, $i=1,\cdots,n$. In the case of the sum of order statistics, we have the following lower bound
\begin{align}
\ell= \bb P \left [ \sum_{i=1}^{\bar{n}}{X^{(i)}(t)} \leq \gamma \right ]  \geq \prod_{i=1}^{n}{\bb P \left [ X_i(t) \leq \frac{\gamma}{\bar{n}}\right ]}.
\end{align}
Note however that for a more complicated choice of $S(\cdot)$, deriving an analytical lower bounds might be not easy and this approach is thus not universally applicable. In such cases, the \emph{inverse complementary CDF} heuristic is preferable, because it can  be applied independently of the choice of $S(\cdot)$. 

\section{Simulation Results}
In this section, we show some numerical results in which we compare, for different scenarios, the proposed multilevel splitting algorithm to some existing estimators.
To perform the comparison, we define the relative error,
i.e. the coefficient of variation using $m$ replications, of an estimator $\hat{\ell}$ as
\begin{align}
\text{RE}(\hat{\ell})=\frac{\sqrt{\var[\hat{\ell}]}}{\hat{\ell}\sqrt{m}}.
\end{align}
We define the work normalized relative variance (WNRV) of an estimator
$\hat{\ell}$ as
\begin{align}
\text{WNRV}(\hat{\ell}) = \text{RE}(\hat{\ell})^2 \times \text{(computing time in seconds)}.
\end{align}
In the following subsections, we consider three important problems: (i) the CDF of weighted sums of Poisson RVs, (ii) the CDF of sums of ordered Weibull and Log-normal RVs, and (iii) the CDF of ratio of independent Log-normal RVs.
For each scenario, we compare the performance of  the multilevel splitting algorithm to some existing estimators.

\subsection{CDF of the Weighted Sum of Independent Poisson RVs}
\begin{table*}[t]
\tiny
\centering
\caption{CDF of the sum of weighted Poisson RVs  with $\lambda_i=1+(i-1)\times 0.2$ and $w_i=i$, $i=1,2,\cdots,12$.}
\begin{tabular}{|c|c|c|c||c|c|c||c|c|c|}
\hline
\multicolumn{1}{|c|}{}& \multicolumn{3}{c||}{Naive MC} & \multicolumn{3}{c|}{Multilevel Splitting} & \multicolumn{3}{c|}{Importance Sampling with $m=6 \times 10^6$} \\
\hline
$\gamma$  & $\hat {\ell}$ & $\re(\hat{\ell}) ($\%$)$  & $\wnrv(\hat{\ell})$ & $\hat {\ell}$ & $\re(\hat{\ell}) ($\%$)$  & $\wnrv(\hat{\ell})$ & $\hat {\ell}$ & $\re(\hat{\ell}) ($\%$)$  & $\wnrv(\hat{\ell})$ \\
\hline
$60$ & $1.02 \times 10^{-4}$  & $ 4.02$ & $0.40$ & $1.07 \times 10^{-4}$  & $1.71$ & $1.62 \times 10^{-2}$ & $1.06 \times 10^{-4}$  & $0.40$ & $3.20 \times 10^{-3}$ \\
\hline
$50$ & $2.00 \times 10^{-5}$  & $ 8.50$ & $1.90$ & $2.26 \times 10^{-5}$  & $1.80$ & $1.82 \times 10^{-2}$ & $2.33 \times 10^{-5}$  & $0.52$ & $5.20 \times 10^{-3}$ \\
\hline
$40$ & $4.83 \times 10^{-6}$  & $ 20.30$ & $10.18$ & $3.97 \times 10^{-6}$  & $2.08$ & $2.69 \times 10^{-2}$ & $3.97 \times 10^{-6}$  & $0.73$ & $1.00 \times 10^{-2}$ \\
\hline
$30$ & $8.33 \times 10^{-7}$  & $ 58.10$ & $84.63$ & $4.80 \times 10^{-7}$  & $2.12$ & $3.75 \times 10^{-2}$ & $5.07 \times 10^{-7}$  & $0.98$ & $1.82 \times 10^{-2}$ \\
\hline
\end{tabular}
\end{table*}

\begin{table*}[t]
\tiny
\centering
\caption{CDF of the sum of order statistics for Weibull Case with $n=8$, $\bar{n}=4$, $\alpha=0.5$, $\eta=1$.}\label{tab2}
\begin{tabular}{|c|c|c|c|c|c|c|c|c|c|}
\hline
\multicolumn{1}{|c|}{} &\multicolumn{3}{c|}{Multilevel splitting } &\multicolumn{3}{c|}{Universal IS estimator with $m=5\times 10^5$} &\multicolumn{3}{c|}{Conditional MC estimator with $m=5\times 10^5$}\\
\hline
$\gamma$ & $\hat {\ell}$ & $RE(\hat{\ell}) \%$& $\wnrv(\hat{\ell})$. &$\hat {\ell}$ & $RE(\hat{\ell}) \%$& $\wnrv(\hat{\ell})$ & $\hat {\ell}$ & $RE(\hat{\ell}) \%$ & $\wnrv(\hat{\ell})$\\
\hline
$1$  &  $0.0029$  &  $0.61$& $4.78 \times 10^{-4}$ &  $0.0029$ & $0.40$  & $7.68 \times 10^{-6}$ & $0.0029$ & $0.12$ &  $1.72 \times 10^{-5}$\\
\hline
$0.5$ &  $3.36 \times 10^{-4}$ & $0.94$    & $1.5 \times 10^{-3}$ & $3.37 \times 10^{-4}$& $0.49$  &  $1.15 \times 10^{-5}$ &  $3.37 \times 10^{-4}$ & $0.13$ & $2.02 \times 10^{-5}$\\
\hline
$0.1$  & $1.26 \times 10^{-6}$  & $1.36$  & $4.5 \times 10^{-3}$ & $ 1.27 \times 10^{-6}$ & $0.66$&  $2.09 \times 10^{-5}$&$1.27 \times 10^{-6}$ & $0.15$ & $2.70 \times 10^{-5}$\\
\hline
$0.05$  & $ 9.80 \times 10^{-8}$ & $1.51$ & $6.4 \times 10^{-3}$ &$9.85 \times 10^{-8}$ & $0.71$  & $2.42 \times 10^{-5}$ &$9.79 \times 10^{-8}$ & $0.16$ & $3.07 \times 10^{-5}$\\
\hline
$0.01$  & $2.10 \times 10^{-10}$ & $1.90$ &$1.43 \times 10^{-2}$ & $2.06 \times 10^{-10}$ & $0.80$ & $3.07 \times 10^{-5}$ & $2.07 \times 10^{-10}$ & $0.17$ & $3.46\times 10^{-5}$\\
\hline
$0.005$ & $1.39 \times 10^{-11}$ & $2.05$ &$2.03 \times 10^{-2}$& $1.39 \times 10^{-11}$& $0.81$ & $3.15 \times 10^{-5}$ & $1.38 \times 10^{-11}$ & $0.17$ & $3.46 \times 10^{-5}$\\
\hline
\end{tabular}
\end{table*}

\begin{table*}[t]
\tiny
\centering
\caption{CDF of the sum of order statistics for Weibull Case with $n=8$, $\bar{n}=4$, $\alpha=0.8$, $\eta=1$.}\label{tab3}
\begin{tabular}{|c|c|c|c|c|c|c|c|c|c|}
\hline
\multicolumn{1}{|c|}{} &\multicolumn{3}{c|}{Multilevel splitting }  &\multicolumn{3}{c|}{Universal IS estimator with $m=5\times 10^5$}  &\multicolumn{3}{c|}{Conditional MC estimator with $m=5\times 10^5$ }\\
\hline
$\gamma$ & $\hat {\ell}$ & $RE(\hat{\ell}) \%$& $\wnrv(\hat{\ell})$. &$\hat {\ell}$ & $RE(\hat{\ell}) \%$& $\wnrv(\hat{\ell})$ & $\hat {\ell}$ & $RE(\hat{\ell}) \%$ & $\wnrv(\hat{\ell})$\\
\hline
$1.03$    & $3.38 \times 10^{-4}$  &  $0.93$ & $1.5 \times 10^{-3}$  & $3.41 \times 10^{-4}$  & $1.28$ & $9.99 \times 10^{-5}$ & $3.37 \times 10^{-4}$  &  $0.1$ & $1.28 \times 10^{-5}$ \\
\hline
$0.38$    & $1.31 \times 10^{-6}$  &  $1.42$ & $5.1 \times 10^{-3}$  & $1.29 \times 10^{-6}$  & $2.31$ & $3.20 \times 10^{-4}$ & $1.31 \times 10^{-6}$  & $0.12$ & $1.74 \times 10^{-5}$ \\
\hline
$0.09$    & $2.09 \times 10^{-10}$ &  $2.00 $ & $1.71 \times 10^{-2}$  & $2.22 \times 10^{-10}$ & $3.20$ & $6.24 \times 10^{-4}$ & $2.10 \times 10^{-10}$ & $0.13$ & $2.06 \times 10^{-5}$ \\
\hline
$0.058$   & $1.36 \times 10^{-11}$ &  $2.29$ & $2.90 \times 10^{-2}$  & $1.33 \times 10^{-11}$ & $3.49$ & $7.43 \times 10^{-4}$ & $1.35 \times 10^{-11}$ & $0.13$ & $2.04 \times 10^{-5}$ \\
\hline
\end{tabular}
\end{table*}

We compare the results given by the proposed multilevel splitting algorithm  to that of the naive MC sampler and the importance sampling estimator that was explained above.
The rates and the weights are selected to be equal to $\lambda_i=1+(i-1)\times 0.2$ and $w_i=i$, $i=1,\cdots,12$.
The number of levels $L$ as well as the splitting times are given by running the second level selection heuristic described above.
The number of samples per level is $s=3000$.
In order to estimate the variance of the multilevel splitting estimator we run the algorithm $m=200$ times and estimate the mean and the variance by sample mean and sample variance respectively.
We use $m=6 \times 10^{6}$ samples in the importance sampling and the naive MC algorithms.

The results in Table I show the inability of the naive MC method to yield a precise estimate of $\ell$ in the rare-event regime. On the other hand, the multilevel splitting and the importance sampling estimators clearly have much better performances than the naive MC sampler. Note however that the values of WNRV reveal that the importance sampling estimator is slightly more efficient than the proposed multilevel splitting estimator (the efficiency is about a factor of 2 when $\gamma=30$).
As we mentioned before, the main advantage of our estimator is its wide applicability. For  a particular application it is possible to construct a more efficient tailor-made  estimator like the one introduced in this article.

\subsection{CDF of the Sum of Ordered RVs}

\begin{table*}[t]
\tiny
\centering
\caption{CDF of the sum of order statistics for Log-normal Case with $n=8$, $\bar{n}=4$, $\mu=0$, $\sigma=2$.}\label{tab6}
\begin{tabular}{|c|c|c|c|c|c|c|c|c|c|}
\hline
\multicolumn{1}{|c|}{}& \multicolumn{3}{c|}{Multilevel splitting} & \multicolumn{3}{c|}{Universal IS estimator with $m=10^6$} &\multicolumn{3}{c|}{Conditional MC estimator with $m=10^6$} \\
\hline
$\gamma$ & $\hat {\ell}$ & $RE(\hat{\ell}) \%$& $\wnrv(\hat{\ell})$. &$\hat {\ell}$ & $RE(\hat{\ell}) \%$& $\wnrv(\hat{\ell})$ & $\hat {\ell}$ & $RE(\hat{\ell}) \%$ & $\wnrv(\hat{\ell})$\\
\hline
$1$     & $8.30 \times 10^{-5}$ & $1.00$ & $2.3 \times 10^{-3}$ &$8.31 \times 10^{-5}$   & $0.68$ & $5.08 \times 10^{-5}$  & $8.31 \times 10^{-5}$   & $0.34$ & $8.57 \times 10^{-4}$\\
\hline
$0.5$   & $1.91 \times 10^{-6}$ & $1.35$ & $5.3 \times 10^{-3}$ & $1.91 \times 10^{-6}$   & $1.27$ & $1.82 \times 10^{-4}$  & $1.90 \times 10^{-6}$   & $0.99$ & $7.2 \times 10^{-3}$\\
\hline
$0.3$   & $7.04 \times 10^{-8}$ & $1.63$ & $1.03 \times 10^{-2}$ & $7.07 \times 10^{-8}$   & $2.11$ & $5.07 \times 10^{-4}$  & $7.00 \times 10^{-8}$   & $2.10$ & $3.19 \times 10^{-2}$\\
\hline
$0.15$  & $3.93 \times 10^{-10}$ & $2.12$ & $2.12 \times 10^{-2}$ &$3.90 \times 10^{-10}$  & $4.37$ & $2.20 \times 10^{-3}$  & $3.92 \times 10^{-10}$  & $5.41$ & $2.09 \times 10^{-1}$\\
\hline
\end{tabular}
\end{table*}

\begin{table*}[t]
\tiny
\centering
\caption{CDF of the sum of order statistics for Log-normal Case with $n=15$, $\bar{n}=15$, $\mu=0$, $\sigma=2$.}\label{tab9}
\begin{tabular}{|c|c|c|c|c|c|c|}
\hline
\multicolumn{1}{|c|}{}& \multicolumn{3}{c|}{Multilevel splitting} & \multicolumn{3}{c|}{Universal IS estimator with $M=5 \times10^7$ }\\
\hline
$\gamma$ & $\hat {\ell}$ & $RE(\hat{\ell}) \%$& $\wnrv(\hat{\ell})$. &$\hat {\ell}$ & $RE(\hat{\ell}) \%$& $\wnrv(\hat{\ell})$ \\
\hline
$3.4$   & $1.93 \times 10^{-6}$  & $1.98$ & $2.48 \times 10^{-2}$ & $2.00 \times 10^{-6}$  & $0.94$ & $7.3 \times 10^{-3}$  \\
\hline
$2.3$   & $6.74 \times 10^{-8}$  & $3.04$ & $6.73 \times 10^{-2}$ &  $6.50 \times 10^{-8}$  & $2.50$ & $5.25 \times 10^{-2}$ \\
\hline
$1.39$   &$3.68 \times 10^{-10}$ & $3.44$ & $9.31 \times 10^{-2}$ &  $4.20 \times 10^{-10}$ & $9.58$ & $6.79 \times 10^{-1}$  \\
\hline
\end{tabular}
\end{table*}
In this part, we show results related to the problem of estimating the CDF of partial sums of ordered RVs. Two distributions are considered: the Weibull and the Log-normal distributions.
Note that the number of samples per level is again $s=3000$ and the algorithm is repeated $m=200$ times in order to be able to estimate the variance. The second heuristic for selecting the levels is again used in this part. The Weibull PDF is given as follows
\begin{align*}
f(x)=\frac{\alpha}{\eta} \left ( \frac{x}{\eta} \right )^{\alpha-1} \exp \left ( -\left ( \frac{x}{\eta}\right )^{\alpha}\right ), \text {    } x\ >0,
\end{align*}
where  $\alpha,\eta>0$ denote the shape and the scale parameters respectively. 
The results in Table II to Table III show the output of a comparison with respect to  two efficient estimators proposed in \cite{8472928} for different values of $n$ and $\bar{n}$. Note that these two estimators achieve the strong criterion of bounded relative error which is generally not achieved by our proposed multilevel splitting estimator. Hence the superior variance reduction of these two estimators in Tables II and  III. However,  when we consider the Log-normal distribution, whose PDF is as follows
\begin{align}
f(x)=\frac{1}{x\sigma\sqrt{2\pi}} \exp \left (- \frac{\left (\log(x)-\mu \right )^2}{2 \sigma^2} \right ), \text{   }  x>0,
\end{align}
 we observe in Tables IV and  V that the proposed estimator is more efficient than these two estimators (the corresponding WNRV is smaller) for particular choices of the system parameters. For example, our proposed estimator is $10$ times more efficient than the conditional MC  estimator for the system parameters of Table IV with $\gamma=0.15$.

\begin{table*}[t]
\tiny
\centering
\caption{CDF of $X_1/(\sum_{i=2}^{n}{X_i}+\eta)$ for Log-normal Case with $n=11$, $\mu_0=20$ dB, $\mu=0$ dB $\sigma=4$ dB, $\sigma_0=6$ dB, $\eta=-10$ dB.}\label{tab10}
\begin{tabular}{|c|c|c|c|c|c|c|}
\hline
\multicolumn{1}{|c|}{}& \multicolumn{3}{c|}{Multilevel splitting} & \multicolumn{3}{c|}{Variance scaling IS with $M=2 \times10^6$ }\\
\hline
$\gamma$ & $\hat {\ell}$ & $RE(\hat{\ell}) \%$& $\wnrv(\hat{\ell})$. &$\hat {\ell}$ & $RE(\hat{\ell}) \%$& $\wnrv(\hat{\ell})$ \\
\hline
$0.02$   & $2.11 \times 10^{-5}$  & $1.41$ & $5.0 \times 10^{-3}$ & $2.01 \times 10^{-5}$  & $2.70$ & $1.33 \times 10^{-2}$  \\
\hline
$0.01$   & $2.30\times 10^{-6}$  & $1.54$ & $6.6 \times 10^{-3}$ &  $2.29 \times 10^{-6}$  & $4.11$ & $3.42 \times 10^{-2}$ \\
\hline
$0.003$   &$2.90 \times 10^{-8}$ & $2.03$ & $1.82 \times 10^{-2}$ &  $2.83 \times 10^{-8}$ & $7.30$ & $1.2 \times 10^{-1}$  \\
\hline
$0.001$   &$2.94 \times 10^{-10}$ & $2.40$ & $3.43 \times 10^{-2}$ &  $3.35 \times 10^{-10}$ & $11.45$ & $1.72 \times 10^{-1}$  \\
\hline
\end{tabular}
\end{table*}

\subsection{CDF of the Ratio of Independent Positive RVs}
We consider the problem of evaluating OP values in the presence of noise and co-channel interferences in a Log-normal fading environment,
i.e. the desired and interfering powers are modeled by Log-normals.
The parameters of the Log-normal distributions are in Table VI.
We compare here the proposed algorithm to that of \cite{7857009}, which is based on observing that the OP is given in this case by the probability that a sum of correlated Log-normal RVs exceeds a given threshold.
An importance sampling approach based on scaling the covariance matrix of the corresponding Gaussian vector is then applied.
The result of the comparison is in Table VI where $m=2 \times 10^6$ samples are used in the importance sampling algorithm of \cite{7857009}.
We use $s=3000$ samples per level and the algorithm is repeated $m=200$ times. Note that the first level selection heuristic is applied here with equally spaced initial intermediate levels $\tilde{t}_{1},\cdots, \tilde{t}_{L}$ with $L=12$.

The values of the relative errors as well as the WNRV show a clear superior performance of the proposed algorithm compared to the importance sampling estimator of \cite{7857009}. For illustration, our proposed algorithm is $5$ times more efficient than the IS estimator when $\gamma=0.001$. Recall also that the proposed estimator is not restricted to the Log-normal fading environment and can be applied regardless of the distributions of the desired and interfering signal powers.
\section{Conclusions}
In this paper, we proposed a dynamic multilevel splitting estimator to efficiently evaluate various time-independent wireless communication systems performances. First, we embed the static or time-independent problem within a continuous-time Markov process so that the distribution of interest corresponds to a snapshot of the Markov process at some time instant. Then, with the Markov process in hand, we applied a version of the classical multilevel splitting to construct our estimator. The resulting splitting algorithm has  wide scope of applicability, as illustrated in our examples of estimating various wireless performance metrics  for a range of common distributions (even discrete distributions). Our numerical experience suggests that the proposed estimator compares favorably with the existing estimators --- sometimes even being the most efficient estimator. 
\bibliographystyle{IEEEtran}
\bibliography{References}
\end{document}

%% file: increasing_splitting.pdf_tex
\begingroup%
  \makeatletter%
  \providecommand\color[2][]{%
    \errmessage{(Inkscape) Color is used for the text in Inkscape, but the package 'color.sty' is not loaded}%
    \renewcommand\color[2][]{}%
  }%
  \providecommand\transparent[1]{%
    \errmessage{(Inkscape) Transparency is used (non-zero) for the text in Inkscape, but the package 'transparent.sty' is not loaded}%
    \renewcommand\transparent[1]{}%
  }%
  \providecommand\rotatebox[2]{#2}%
  \newcommand*\fsize{\dimexpr\f@size pt\relax}%
  \newcommand*\lineheight[1]{\fontsize{\fsize}{#1\fsize}\selectfont}%
  \ifx\svgwidth\undefined%
    \setlength{\unitlength}{347.33189839bp}%
    \ifx\svgscale\undefined%
      \relax%
    \else%
      \setlength{\unitlength}{\unitlength * \real{\svgscale}}%
    \fi%
  \else%
    \setlength{\unitlength}{\svgwidth}%
  \fi%
  \global\let\svgwidth\undefined%
  \global\let\svgscale\undefined%
  \makeatother%
  \begin{picture}(1,0.57493452)%
    \lineheight{1}%
    \setlength\tabcolsep{0pt}%
    \put(0,0){\includegraphics[width=\unitlength,page=1]{increasing_splitting.pdf}}%
    \put(0.07438436,0.00942132){\color[rgb]{0,0,0}\makebox(0,0)[lt]{\lineheight{1.25}\smash{\begin{tabular}[t]{l}0\end{tabular}}}}%
    \put(0.89718151,0.00972497){\color[rgb]{0,0,0}\makebox(0,0)[lt]{\lineheight{1.25}\smash{\begin{tabular}[t]{l}1\end{tabular}}}}%
    \put(0.45743369,0.00901645){\color[rgb]{0,0,0}\makebox(0,0)[lt]{\lineheight{1.25}\smash{\begin{tabular}[t]{l}$t_1$\end{tabular}}}}%
    \put(0.68897911,0.00901645){\color[rgb]{0,0,0}\makebox(0,0)[lt]{\lineheight{1.25}\smash{\begin{tabular}[t]{l}$t_2$\end{tabular}}}}%
    \put(-0.0019119,0.55743505){\color[rgb]{0,0,0}\makebox(0,0)[lt]{\lineheight{1.25}\smash{\begin{tabular}[t]{l}$S(\boldsymbol{X}(t))$\end{tabular}}}}%
    \put(0.04862152,0.29845131){\color[rgb]{0,0,0}\makebox(0,0)[lt]{\lineheight{1.25}\smash{\begin{tabular}[t]{l}$\gamma$\end{tabular}}}}%
    \put(0,0){\includegraphics[width=\unitlength,page=2]{increasing_splitting.pdf}}%
  \end{picture}%
\endgroup%